# Liquid Hole-Multipliers:
# A potential concept for large single-phase noble-liquid TPCs of rare events

Amos Breskin

*Department of Astrophysics and Particle Physics*
*Weizmann Institute of Science*
*76100 Rehovot, Israel*

## Abstract

A novel concept is proposed for large-volume single-phase noble-liquid TPC detectors for rare events. Both radiation-induced scintillation-light and ionization-charge are detected by Liquid Hole-Multipliers (LHM), immersed in the noble liquid. The latter may consist of cascaded Gas Electron Multipliers (GEM), Thick Gas Electron Multiplier (THGEM) electrodes or others, coated with CsI UV-photocathodes. Electrons, photo-induced on CsI by primary scintillation in the noble liquid, and event-correlated drifting ionization electrons are amplified in the cascaded elements primarily through electroluminescence, and possibly through additional moderate avalanche, occurring within the holes. The resulting charge-signals or light-pulses are recorded on anode pads or with photosensors – e.g. gaseous photomultipliers (GPM), respectively. Potential affordable solutions are proposed for multi-ton dark-matter detectors; open questions are formulated for validating this dream.

Amos.breskin@weizmann.ac.il

As a result of a dream, eventually based on solid experimental facts, we propose and discuss a new possible concept of a *single-phase* noble-liquid Time Projection Chamber (TPC) for large-volume detectors of rare events.

Noble-liquid detectors have been employed since a few decades in particle-physics calorimetry, gamma imaging in astronomy and medical fields (e.g. PET, Compton Camera), as well as in astroparticle physics. In the latter, most important are neutrino physics and dark-matter (DM) searches. Detailed surveys of present techniques and their applications can be found in recent reviews [1, 2, 3].

One of the most advanced methods of direct detection of DM signatures, in form of Weakly Ionizing Massive Particles (WIMPs), is the observation of nuclear-recoil rates they deposit in rare scattering events in a noble-liquid TPC detector. The extremely low interaction cross sections involved require large-mass target-detectors, with the main challenges being the detection of faint low-energy (keV-scale) signals and effective background suppression. Liquid argon (LAr, e.g. ArDM [4]) and liquid xenon (LXe, e.g. XENON100 [5], LUX [6]) are the preferred targets, allowing for the construction of large volumes of homogeneous detection media. They operate either in a *single-phase* (liquid), or *dual-phase* (liquid and gas) configurations. *Single-phase* detectors (e.g. XMASS LXe experiment [7]) are simpler, relying on measuring the scintillation light emitted promptly at the site of interaction. *Dual-phase* detectors operate as TPCs, recording two signals: the primary prompt scintillation light (S1) within the liquid and a secondary delayed signal (S2) generated by the recoil-induced ionization electrons liberated at the interaction site, as they pass through the gas phase after extraction from the liquid. S2 can be a result of electroluminescence in the saturated vapor phase above the liquid [5] or of a multiplied charge [4] in this gas gap. Background suppression is achieved by running the experiments in underground facilities, using radio-pure detector materials and applying passive shielding and active veto schemes. The distinct difference of the secondary-to-primary signals ratio (S2/S1), in *dual-phase* detectors, for nuclear recoils and for electron recoils from gamma background is the key to their efficient discrimination as demonstrated in [5].

The sensitivity of DM detectors to detect low-energy WIMP-induced recoils and effectively discriminate them from background depends largely on the photon detectors. Present-day DM detectors and others under conception (e.g. XENON1ton [8]) employ large and very costly vacuum-photomultiplier (PMT) arrays; these must withstand cryogenic conditions and have low natural radioactivity and high single-photon detection efficiency at the relevant UV-emission wavelengths. While present PMTs reasonably fulfill the strict requirements of current experiments, novel affordable solutions are required for future generations of multi-ton detectors, e.g. DARWIN [9]. For such experiments, the price of present-type photon detectors would become exorbitant – calling for new solutions.

One concept, under advanced R&D, relies on photon recording in *dual-phase* TPCs with large-area gas-avalanche photomultipliers (GPMs) [10, 11, 12]. These are expected to have lower cost, flexible experiment-adapted flat geometry, large size, high pixilation and potentially low background. A typical GPM consists of cascaded hole avalanche multipliers (e.g. Gas Electron Multipliers (GEM[13]), Thick Gas Electron Multipliers (THGEM [14])) or hybrid combinations of hole and mesh multipliers [15, 16], with the first element being coated with a robust CsI UV-photocathode [17]. Both the THGEM-GPM and the hybrid-GPM showed high gains in combination with a LXe-TPC [15, 16]. The GPM concept has been under intense R&D at the Weizmann Institute, within DARWIN [9], for future multi-ton DM detectors; it has been considered by the PANDA projected DM experiment, also in a *single-phase* configuration [18]. A *single-phase* LXe detector coupled to a GPM is under advanced R&D, for combined fast-neutron and gamma imaging [19].

While *dual-phase* noble-liquid detectors have become rather "standard" instruments in present experiments, the expansion of their geometrical dimensions in multi-ton devices might not be technically simple; particularly because of the necessity to extract electrons from liquid into gas through large, very flat mesh-electrodes, keeping constant temperature (and pressure) conditions across the detector. We therefore propose a novel challenging concept that would permit the efficient recording of both low-energy recoil-induced scintillation-light and ionization-electron signals in large-volume *single-phase* noble-liquid detectors.

In our new concept, both radiation-deposited scintillation light and ionization charges are collected within the liquid into novel Liquid Hole-Multipliers (LHM), immersed within the liquid. The LHMs consist of a combination of cascaded GEMs, THGEMs, Micro-Hole & Strip Plates (MHSPs) [20] or other dedicated electrodes, coated with CsI photocathodes; the latter have high quantum efficiency values for both LAr and LXe emission wavelengths [16], also in liquid xenon [21, 22]. The process is schematically shown in Figure 1. Electrons, photo-induced on CsI by primary scintillation and event-correlated drifting ionization electrons are collected within the liquid, by a strong dipole field, into the multiplier's amplification holes. They generate electroluminescence within the holes under high electric fields at their center, and, according to the field-strength, they may also undergo some modest charge multiplication. Forward-emitted UV-photons (shown in Figure 1) impinge on the photocathode of the next element in the cascade, inducing additional photoelectrons that generate UV photons in its holes; the process can be repeated until the right sensitivity is reached (even without charge multiplication). According to the total amplification reached, the final charge signals can be collected on readout pads; alternatively, photon flashes from the last element can be recorded by photon detectors deployed within the liquid – preferably GPMs (Figure 1). To avoid possible photon-feedback effects generated by electroluminescence (e.g. secondary electrons from electroluminescence-photons emitted backwards) the hole-electrodes of the cascade could be staggered.

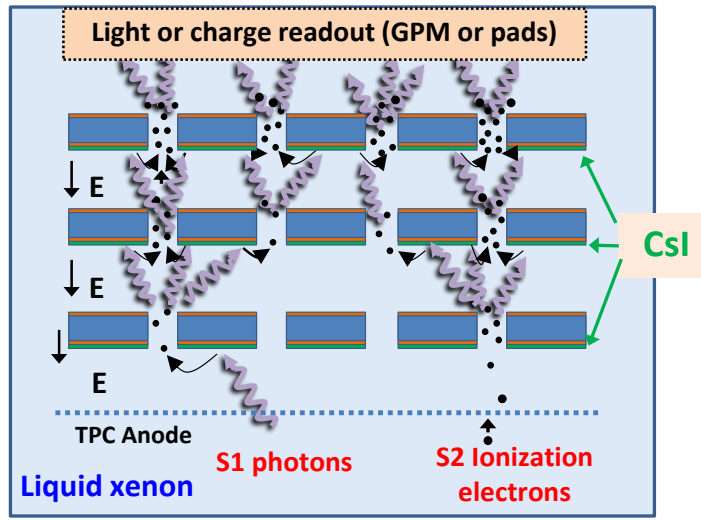

*Fig. 1 The proposed Liquid Hole-Multiplier (LHM) principle. Light-amplification and optional modest charge multiplication in sensors immersed within the noble liquid permits detection of both radiation-induced scintillation UV-photons (S1) and ionization electrons (S2). Shown is an example with 3 THGEM electrodes:*
- *Radiation-induced UV-photons impinge on the first CsI-coated THGEM electrode ;*
- *Extracted photoelectrons are trapped into the holes, where high fields induce electroluminescence (and possibly small charge gain); The resulting forward-emitted photons (the only ones shown here) are further amplified by a cascade of CsI-coated THGEMs .*
- *Similarly, drifting radiation-induced ionization electrons are focused into the holes and follow the same amplification path .*
- *S1 and S2 signals are recorded optically by an immersed Gaseous Photomultiplier (GPM) or by charge collected on pads.*

It should be noted that the working hypotheses for this new concept have been derived from the following known facts:

- The existing knowledge that moderate charge multiplication (a few hundreds at best) [23, 24] and electroluminescence (~100 photons/electron at gain ~50) [25] are possible on a few-microns diameter anode wires immersed in LXe, as also reviewed in [26, 3]. To our best knowledge, no stable gain was reported on thin wires in LAr; charge gains of ~100 were reached only on sharp (0.25 radius) tips in LAr [27];
- A recent demonstration that electroluminescence (estimated yields reaching 500 photons/electron), without charge multiplication, occurs in THGEM holes in LAr, at rather moderate fields [28, 12];
- Similar photon-mediated amplification concepts, in a cascaded detector, were already demonstrated by our research team (Weizmann/Aveiro/Coimbra) [29] and in parallel by others [30], in a gas phase, with the aim of blocking avalanche ions.
- CsI photocathodes are known to have high quantum yields in LXe (~23% at 178nm under 10kV/cm) [21, 22], which was the basis for some potential *dual-phase* TPCs [22, 31] and *single-phase* spherical-TPC [32] noble-liquid detector concepts (not yet materialized). The latter, of a spherical geometry, was conceived with charge readout after multiplication within the liquid in a central spherical GEM detector.

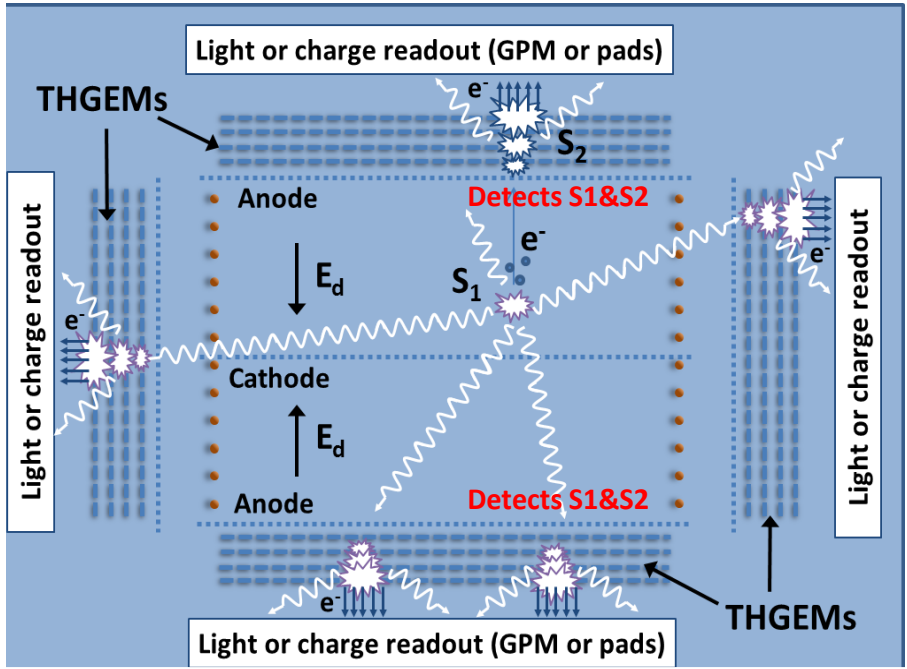

*Fig. 2 A proposed dual-sided single-phase TPC DM detector with top, bottom and side THGEM-LHMs (or other LHMs). The prompt S1 scintillation signals are detected with all LHMs. The delayed S2 ionization signals are recorded with the bottom and top LHMs. The S1 and S2 signals are recorded optically by an immersed GPM or by charges collected on pads.*

As the LHM is sensitive to both electrons and photons (Figure 1), one could conceive *single-phase* symmetric noble-liquid detectors, with bi-directional drift volumes, having additionally photosensors at their circumference (Figure 2). The net advantages would be: a two-fold reduced drift potential (with significant technical benefits) and a very efficient primary-scintillation signal collection over a broad solid angle – with a potentially lower detection threshold for low-mass WIMPs. A further embodiment (Figure 3) would be a large volume composed of a stack of further shorter bi-directional drift volumes with multiple-LHM elements read out by pads of flat GPMs.

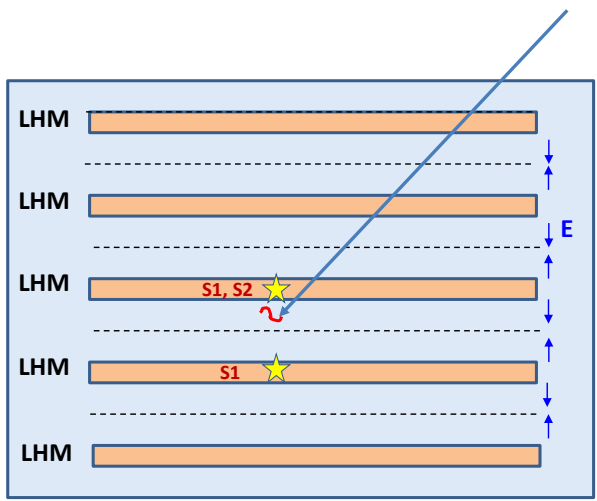

*Fig. 3 A possible layout of a large-volume noble-liquid DM detector composed of a stack of short, bi-directional, drift gaps with multiple double-sided LHM elements. The S1 and S2 signals induced by an event are recorded by pads of by flat GPMs. Drawing not to scale.*

The validation of this challenging *dream-concept* requires crossing many potential pitfalls. The studies are in course in the novel LXe cryostat system (WILiX) assembled at the Weizmann Institute; they necessitate careful investigations of numerous relevant issues and parameters. A success will rely first of all on the possibility of reaching sufficient light amplification, and maybe small charge multiplication, in liquid phase, in single- and cascaded-multipliers and validating the photon-assisted multiplication in liquid phase [29]. Unlike single-wire structures [26], cascaded elements have the potential of reaching high total gains, with a moderate gain per single element, e.g. as shown in [16]. While light amplification onset in THGEM holes was measured in LAr at relatively low fields at the hole center (~60kV/cm) [28, 12], some literature results [26] indicate upon higher field values, of ~100 and 1000kV/cm necessary for the respective onset of scintillation and charge multiplication on thin wires in LXe. If necessary, the hole-multipliers could be interlaced with thin amplification-wires; such hybrid multipliers would provide higher total light and charge yields due to the high reachable fields at the wire's vicinity at comparatively low potentials, as summarized in [26]. The multipliers' geometry must be optimized also for high collection efficiency of both ionization electrons and photoelectrons into the holes, e.g. similarly to [33]. Other issues would be the choice of high-purity electrode materials (at a later stage, also radio-pure ones), understanding the role of impurities in the liquid, photon-feedback suppression, stability, evaluating pads vs. optical readout etc. Finally, one should evaluate the suitability of the detector's pulse-height and time resolutions, detection thresholds and background suppression capabilities for rare-event detection.

The validation of the new proposed concept could have considerable impact on the conception of the next generation of large-mass LXe or LAr detectors of DM and other rare events. The potential highlights would be: a rather simple and probably economic construction of PMT-free detectors sensitive to both scintillation light (S1) and ionization electrons (S2), larger scintillation signals (very large solid angle) with resulting lower expected detection thresholds and shorter drift lengths (Figs 2, 3) resulting in lower drift voltages and lower electron losses. If successful, the new LHM concept could pave ways towards other fields of applications.

I would like to thank Dr. Lior Arazi and Dr. Rachel Chechik of the Weizmann Institute and Prof. Elena Aprile of Columbia University for helpful discussions and collaboration in this project. This work was supported in part by the Israel Science Foundation (Grant 477/10) and the MINERVA Foundation (Project 710827). A. Breskin is the W.P. Reuther Professor of Research in the Peaceful use of Atomic Energy.